\documentclass[aps,prc,preprint]{revtex4}
 \usepackage{amsfonts}
\usepackage{amssymb}
\usepackage{graphics}
\usepackage{graphicx}
\usepackage{color}
\begin{document}

\title{Coexistence of phases in asymmetric nuclear matter under strong magnetic fields}

\author{R. Aguirre$^{1,2}$ and E. Bauer$^{2,3}$}
\email{aguirre@fisica.unlp.edu.ar, bauer@fisica.unlp.edu.ar}
\affiliation{$^1$Departamento de Fisica, Facultad de Ciencias
Exactas, Universidad Nacional de La Plata, Argentina}
\affiliation{$^2$ IFLP, CCT-La Plata, CONICET. Argentina}
\affiliation{$^3$Facultad de Ciencias Astron\'omicas y
Geof\'{\i}sicas, Universidad Nacional de La Plata, Argentina}

\begin{abstract}
The equation of state of nuclear matter is strongly affected by
the presence of a magnetic field. Here we study the equilibrium
configuration of asymmetric nuclear matter for a wide range of
densities, isospin composition, temperatures and magnetic fields.
Special attention is paid to the low density and low temperature
domain, where a thermodynamical instability exists. Neglecting
fluctuations of the Coulomb force, a coexistence of phases is
found under such conditions, even for extreme magnetic
intensities. We describe the nuclear interaction by using the
non--relativistic Skyrme potential model within a Hartree--Fock
approach. We found that the coexistence of phases modifies the
equilibrium configuration, masking most of the manifestations of
the spin polarized matter. However, the compressibility and the
magnetic susceptibility show clear signals of this fact. Thermal
effects are significative for both quantities, mainly out of the
coexistence region.
\end{abstract}


\maketitle

\section{Introduction}

The dense nuclear matter under magnetic fields has been
intensively studied (see~\cite{LAI} and references therein),
particularly in relation to astrophysical issues. Investigations
of the neutron star structure~\cite{LATTIMER} as well as the
cooling of magnetized stars~\cite{SHIBANOV, YAKOVLEV, BEZCHASTNOV}
need the equation of state for magnetized matter as an important
input. The presence of very intense magnetic fields in compact
stellar objects has been proposed, based on the observational
evidence of periodic or irregular radiation from localized
sources. According to the energy released and the periodicity of
the episodes, these objects have been classified as pulsars, soft
gamma ray repeaters and anomalous X-ray pulsars. They have been
associated with different stages of the evolution of neutron
stars. On the star surface the magnetic field could reach values
$10^{14}-10^{15}$ G, as in the case of magnetars and it is
expected a growth of several orders of magnitude in its dense
interior.

Recent investigations \cite{KHARZEEV,MO,SKOKOV} have pointed out
that matter created in heavy ion collisions could be subject to
very strong magnetic fields. As a consequence the particle
production would exhibit a distinguishable anisotropy. A
preferential emission of charged particles along the direction of
the magnetic field is predicted in~\cite{KHARZEEV} for noncentral
heavy ion collisions, due to magnetic intensities  $e\, B \sim
10^2$ MeV. Improved calculations taking care of the mass
distribution of the colliding ions \cite{MO}, does not modify
essentially the magnitude of the produced fields.  Furthermore,
the numerical simulations performed by \cite{SKOKOV} predict
larger values $e\, B \sim m_\pi^2 \sim 2 \times 10^4$ MeV$^2$.

The effects of magnetic fields on a dense nuclear environment have
been described using different models  \cite{CHAKRABARTY,
BRODERICK, YUE2,SINHA,
RABHI,RYU,RYU2,ANG,DONG,DIENER,DONG2,PGARCIA,
ISAYEV,UNLP,UNLP2,BORDBAR,BIGDELI2}. For instance, covariant field
theoretical models have been used to study the role of the
magnetic field on hyperonic matter \cite{YUE2,SINHA},
instabilities at subsaturation densities \cite{RABHI,RYU},
magnetization of stellar matter \cite{DONG}, saturation properties
of symmetric matter \cite{DIENER} and the symmetry energy
\cite{DONG2}. Non-relativistic models have also been used, in the
effective interactions of  \cite{PGARCIA, ISAYEV,UNLP} or the
microscopic models used in the variational calculations of
\cite{BORDBAR, BIGDELI2}. A comparison of neutron matter results,
using different models was presented in \cite{UNLP2}.

It is a well known fact that the nuclear environment experiences
thermodynamical instabilities at subnuclear densities and low
temperatures. Evidence of this phenomenon can be found in the
isospin distillation effect for heavy ion collisions~\cite{DAS}. These
instabilities give rise to a coexistence of phases if the surface
tension is low enough. A more complex scenario is obtained when an
external magnetic field is added, since there is a competition
among two opposite trends. On one hand the magnetic force induces
a globally ordered state with aligned spins. On the other hand the
nuclear interaction favors the coexistence of phases where two
states of different densities and spin polarizations are combined
in order to lower the free energy.

In the present work we explore the possibility of a
coexistence of phases for nuclear matter under strong magnetic
fields, taking as  variables the nuclear density, the isospin
composition of matter, and the temperature $T < 10$ MeV.
The possible fields of applications, such as
those mentioned before, show a complex scenario where the detailed
physical mechanisms are not easy to understand because there is a
superposition of effects which can combine to give very different
manifestations. Therefore, we aim to present here some of the
variables appearing in realistic situations, emphasizing the role
of each of these factors, and to understand how they interact in a
specific environment. Special attention is paid to the relevant
quantities associated with them, as the spin polarization, the
isothermal compressibility and the magnetic susceptibility. With
this purpose in mind we have analyzed a wide range of isospin
composition and we have also reached the extreme value $B=10^{19}$
G for the external magnetic field. This selection emphasizes the
effects of these variables, which under certain conditions can
appear weakened, or hidden by another factors.

The Skyrme model \cite{SKYRME, DOUCHIN, BENDER1, BENDER2} is
appropriate to describe the nuclear interaction under the
conditions of interest. This is a non-relativistic effective
model where the in--medium nuclear force is simulated by a density
dependent potential. It was successfully used to describe atomic
nuclei as well as nuclear matter properties.

This article is organized as follows. We review the Skyrme model
for nuclear matter under an external magnetic field in the next
section, a brief resume of the Gibbs construction for the
coexistence of phases is presented in Section \ref{SecIII}, the
results are shown and discussed in Section \ref{SecIV}. A final
summary and the main conclusions are given in Section
\ref{Summary}.

\section{Spin polarized nuclear matter in the Skyrme model}\label{SecII}

The Skyrme model is an effective formulation of the nuclear
interaction which has been employed profusely in the literature
\cite{BENDER1}. It consists of a basic Hamiltonian with contact
nucleon-nucleon potentials including density dependent
coefficients,
\begin{eqnarray}
v_{Sky}(r_1,r_2)&=&t_0 (1+x_0 \, P_\sigma)
\delta(r_1-r_2)+\frac{1}{2}t_1 (1+x_1 \, P_\sigma)
\left[\stackrel{\leftarrow}{q}^2
\delta(r_1-r_2)+\stackrel{\rightarrow}{q}^2
\delta(r_1-r_2)\right] \nonumber\\
&+&t_2 (1+x_2 \, P_\sigma)\stackrel{\leftarrow}{q} \cdot
\delta(r_1-r_2) \stackrel{\rightarrow}{q}+\frac{1}{6}t_3 (1+x_3 \,
P_\sigma) \delta(r_1-r_2) n^\sigma((r_1+r_2)/2) \nonumber \\
&+& i W_0 (\sigma_1+\sigma_2)\cdot \stackrel{\leftarrow}{q} \times
\delta(r_1-r_2) \stackrel{\rightarrow}{q}\nonumber
\end{eqnarray}
where $\sigma_k$ represent the Pauli matrices for spin,
$P_\sigma=(1+\sigma_1 \cdot \sigma_2)/2$ is the spin exchange
operator, $q=-i (\nabla_1 - \nabla_2)/2$ is the relative momentum
operator and $n$ is the total baryonic density.\\
Note that throughout this article we use units such that $c=1$,
$\hbar =1$.\\
 The
interaction--parameters are fixed to cover a variety of
applications such as exotic nuclei or stellar matter. Using the
Hartree-Fock approximation, one can find an energy density
functional, which is a convenient way to study thermodynamical
properties of the system.

We are particularly interested in the contributions coming from
terms containing time reversal-odd densities and currents, since
they are active when spin states are not symmetrically occupied. A
derivation of these terms  can be found in \cite{BENDER2}. We
assume the magnetic field has not dynamics, that is, it behaves as
an external field. There is a direct coupling between nucleons and
the magnetic field, due to their intrinsic magnetic moments. This
implies an additional term $-\mu_N \chi_a B$ to the single
particle spectra of the standard Skyrme model, where $\mu_N$ is
the Bohr magneton and the Lande factors $\chi_a$ take account of
the anomalous magnetic moments. They take the values
$\chi_1=2.793$ and $\chi_2=-1.913$ for protons and
neutrons, respectively.\\
Furthermore, the magnetic field induce a quantization of the
energy spectra of charged particles \cite{LANDAU}. In the case of
a uniform field, the corresponding Schr\"{o}dinger equation
exhibits quantized eigenvalues, associated with the motion in
directions orthogonal to the applied field. They are
oscillator-like levels, depending on a discrete quantum number in
the form $(j+1/2)\, \omega$, with $\omega = e B/m$ the cyclotron
frequency of the particle. We can summarize the effects of a
uniform magnetic field over the spectra of nucleons by
\begin{eqnarray}
\varepsilon_{1sj}(p_z)&=&\frac{p_z^2}{2 m^*_{1s}}+\frac{1}{8}
v_{1s} + \mu_N B (2 j + 1 - s \chi_1),
\label{protSP} \\
\varepsilon_{2s}(p)&=&\frac{p^2}{2 m^*_{2s}}+\frac{1}{8} v_{2s}-
\mu_N B s \chi_2. \label{neutSP}
\end{eqnarray}
The first two terms  in the {\it r.h.s} of these equations, are
the common results for the Skyrme model, which now have an
implicit dependence on the field $B$. The spin index $s=1$
($s=-1$) denotes a spin--up (spin--down) projection, the effective
nucleon mass $m^*_{as}$ is defined by
\begin{equation}
\frac{1}{m^*_{as}}=\frac{1}{m}+\frac{1}{4}\, n\,(b_0-b_2 w I_a)
+\frac{1}{4}\, s \sum_b (b_1 + I_a I_b b_3)\; W_b \label{EffMass}
\end{equation}
with $m$ the degenerate nucleon mass in vacuum, $w=(n_2-n_1)/n$ is
the isospin asymmetry fraction, with $n_1, \, n_2$ standing for
the particle number density of protons and neutrons respectively.
Note that $n=n_1+n_2$. Since the spin states are not symmetrically
occupied, one can define for each isotopic component the number
density of particles with a given spin polarization $n_{a\,s}$.
The spin asymmetry density $W_a$ gives a measure of the spin
polarization  $W_a=\sum_s s \, n_{a\,s}$, clearly $n_a=\sum_s n_{a
\, s}$. We have defined $I_a=1, \; (-1)$ for
protons (neutrons).\\
In Eqs. (\ref{protSP}) and (\ref{neutSP}) we have used the single
particle Skyrme potential energy
\begin{eqnarray}
v_{as}&=&(a_0 - a_2 w I_a)\, n + \sum_{s' c} (b_0 +I_a I_c b_2)
K_{c s'} + s \sum_c (a_1 + a_3 I_a I_c) W_c +
\nonumber \\
&& + s \sum_{s' c} s' (b_1 + b_3 I_a I_c) K_{c s'}.
\label{SkmPotential}
\end{eqnarray}
The expressions for the kinetic energy density  $K_{as}$ will be
presented below. The Eqs. (\ref{protSP})--(\ref{SkmPotential})
have been written in terms of a set of density dependent
coefficients $a_0-a_3$ and $b_0 - b_3$, which are related to the
standard parameters of the Skyrme model by,
\begin{eqnarray}
a_0& = &6 t_0 + t_3 n^\sigma, \nonumber \\
b_0&= &[3 t_1 + t_2 (5 + 4 x_2)]/2 \nonumber \\
a_1&= &-2 t_0 (1 - 2 x_0)- t_3 (1 - 2 x_3) n^\sigma/3 \nonumber \\
b_1&= &[t_2 (1 + 2 x_2) - t_1(1 - 2 x_1)]/2 \nonumber\\
a_2&= &-2 t_0 (1 + 2 x_0)- t_3 (1 + 2 x_3) n^\sigma/3 \nonumber \\
b_2&= &[t_2 (1 + 2 x_2) - t_1(1 + 2 x_1)]/2 \nonumber\\
a_3&= &-2 t_0- t_3 n^\sigma/3 \nonumber \\
b_3&= &(t_2 - t_1)/2 \nonumber
\end{eqnarray}
We assume baryonic and isospin number conservation, therefore
independent chemical potential $\mu_a$ can be assigned to protons
and neutrons. The corresponding distribution functions
$f_{as}(T,p)=\left[1+\exp (\varepsilon_{as}(p)-\mu_a)/T
\right]^{-1} $ are the Fermi occupation number for a particle at
temperature $T$, with momentum \mbox{\boldmath$p$} and isospin and
spin projections  $a$ and $s$, respectively.

Now we show explicit expressions for the density of particles with
a given spin polarization, the kinetic energy and the isospin
asymmetry densities, separately for protons and neutrons
\begin{eqnarray}
n_{1s}&=& \frac{e B}{(2 \pi)^2}\sum_j \int_{-\infty}^\infty dp_z
\;
f_{1sj}(T,p_z ) \\
W_1&=&\frac{e B}{(2 \pi)^2}\sum_{s,j} s \int_{-\infty}^\infty
dp_z \; f_{1sj}(T,p_z)\\
K_{1s}&=&\frac{e B}{(2 \pi)^2}\sum_j \int_{-\infty}^\infty dp_z
\;p_z^2 f_{1sj}(T,p_z)\\
n_{2s}&=& \frac{1}{(2 \pi)^3}\int d^3p \; f_{2s}(T,p)\\
W_2&=&\sum_s  \frac{s}{(2 \pi)^3}\int d^3p \; f_{2s}(T,p)\\
K_{2s}&=&\frac{1}{(2 \pi)^3}\int d^3p \;p^2 f_{2s}(T,p)
\end{eqnarray}

For the proton related quantities, we have taken into account
that, assuming \mbox{\boldmath$B$} along the z-axis, each
eigenstate spreads over a bounded region of area $2 \pi e B$ in
the $p_x-p_y$ plane. The component $p_z$ is not bounded and varies
continuously. Therefore, the contribution of a charged particle to
macroscopic quantities per unit volume has been evaluated by means
of the  replacement $ \int d^3p/(2 \pi)^3 \rightarrow e B\int
dp_z/(2 \pi)^2$.

The energy density can be split into two terms,
\begin{equation}
\mathcal{E}=\mathcal{E}_{Skm} + \mu_N B \left(2 L + n_1 - \chi_1
W_1 - \chi_2 W_2 \right), \label{EnergiaTotal}
\end{equation}
one of them depends on $B$. The remaining one is similar to the
common contribution of the Skyrme model ${\mathcal E}_{Skm}$ in a
Hartree-Fock approach, but now it depends implicitly on the
magnetic intensity

\begin{equation}
 {\mathcal E}_{Skm}=\sum_{a,s}\, \frac{K_{as}}{2
m^*_{as}}+\frac{1}{16}\left[a_1 \left( \sum_a W_a\right)^2 + a_3
\left( \sum_a I_a W_a\right)^2 + (a_0+a_2 w^2)\; n^2\right].
\label{Enerdens}
\end{equation}
In Eq. (\ref{EnergiaTotal}) we used,
\begin{equation}
L=\frac{e B}{(2 \pi)^2} \sum_{s, j} j\int_{-\infty}^\infty dp_z
\; f_{1sj}(T,p_z ).
\label{Lterm}
\end{equation}

For completeness we also show the expression for the entropy
density ${\cal S}$ in the quasi-particle approach,
\begin{eqnarray}
{\cal S}&=&{\cal S}_1+{\cal S}_2, \nonumber \\
{\cal S}_1&=&-  \frac{e B}{(2 \pi)^2}\sum_{s,j}
\int_{-\infty}^\infty dp_z \;\left[ f_{1sj}\ln\left(f_{1sj}\right) +
\left(1- f_{1sj}\right) \,\ln\left(1- f_{1sj}\right)\right],\nonumber \\
{\cal S}_2&=&- \sum_s\frac{1}{(2 \pi)^3}\int d^3p \;\left[
f_{2s}\ln\left(f_{2s}\right) + \left(1- f_{2s}\right)
\,\ln\left(1- f_{2s}\right)\right] \nonumber
\end{eqnarray}
The entropy is needed to build up the
free energy $\mathcal{F}=\mathcal{E}-T
\mathcal{S}$ and the pressure $P=\sum_{a} \mu_a n_a -\mathcal{F}$.
The magnetization of the system $\mathcal{M}$ is evaluated in
terms of the grand canonical potential $\Omega(\mu_k,T,V)$
according to \cite{PATHRIA},

\begin{equation}
\mathcal{M}= \frac{1}{V}\left( \frac{\partial \Omega}{\partial
B}\right)_{\mu,T,V}.
\end{equation}

For the system considered, we have $\Omega=-P\, V$. As expected,
it can be decomposed into proton and neutron contributions
$\mathcal{M}=\mathcal{M}_1+\mathcal{M}_2$. Finally, the standard
relations are used for the isothermal compressibility $K$ and
magnetic susceptibility $\chi$,
\begin{eqnarray}
K&=&-\frac{1}{V}\left(\frac{\partial V}{\partial P}
\right)_{N_1,N_2,T,B} \nonumber \\
\chi_a &=& \left(\frac{\partial \mathcal {M}_a}{\partial
B}\right)_{N,T,V} \nonumber
\end{eqnarray}
Note that in our scheme we were able to develop analytical expressions for
the isothermal compressibility and magnetic susceptibility. This gives us some
confidence in the evaluation of these magnitudes, as for instance, the
susceptibility for low temperatures has fast variations with the
density.

In our approach both proton and neutron numbers are conserved
separately, therefore the states of polarization of each component
are also independent. The global polarization is determined by the
condition of minimum free energy $F$.  This criterium differs from
that in \cite{UNLP}, where the Legendre transformed potential
$\mathcal{F - M} B$ was used.

The equilibrium state has a variable spin configuration, depending
on $n, \,w, \, T$ and $B$.  As will be shown, in the low
temperature and low density domain the coexistence of phases
imposes a state with a lower degree of polarization than the case
which does not consider the phase transition.

\section{Coexistence of phases in nuclear matter}\label{SecIII}

The nuclear interaction gives rise to instabilities in a dense
infinite medium at low temperatures. If the Coulomb interaction is
taken into account and its fluctuations are included, a nucleation
process can be found.\\
Under the hypothesis assumed in the present work, the system
evolves through a succession of equilibrium states, where it
decomposes spontaneously into two phases of different density and
isospin composition. This phenomenon has been classified as a
non--congruent phase transition \cite{SCHRAMM} since there are two
conserved charges, i.e. proton and neutron numbers. Its importance
in the study of the in--medium nuclear interaction has been
emphasized in recent investigations \cite{WEISE}.

These two coexisting phases, distinguished in the following by
superindices $a$ and $b$, have different numbers of proton and
neutrons. However they are subject to the thermodynamical
equilibrium conditions,
\begin{eqnarray}
P(N_1^a,N_2^a, T,V^a,B)&=&P(N_1^b,N_2^b, T,V^b,B), \label{PresCR} \\
\mu_1(N_1^a,N_2^a, T,V^a,B)&=&\mu_1(N_1^b,N_2^b, T,V^b,B),\label{PQ1CR} \\
\mu_2(N_1^a,N_2^a, T,V^a,B)&=&\mu_2(N_1^b,N_2^b,
T,V^b,B), \label{PQ2CR}
\end{eqnarray}
Furthermore, each phase contributes to every intensive additive
physical quantity, such as the densities of energy, entropy, etc.
Thus, the free energy per unit volume and the density number of
nucleons for the whole system can be written as
\begin{eqnarray}
\mathcal{F}(N_1,N_2,T,V,B)&=&(1-\lambda)\,
\mathcal{F}(N_1^a,N_2^a,T,V^a,B)+ \lambda \,
\mathcal{F}(N_1^b,N_2^b,T,V^b,B),\nonumber \\
n_k&=&(1-\lambda)\,n_k^a+ \lambda \, n_k^b, \;\; k=1,2 \nonumber
\end{eqnarray}
where the coefficient $\lambda$ can be interpreted as the fraction
of the partial volume occupied by the state $b$, hence it is
bounded by $0\leq \lambda \leq 1$.

In Fig. 1 it is shown how this procedure, commonly known as the
Gibbs construction, works for the pressure and the spin asymmetry
quotient of each component $W_k/n_k$. The general features of this
figure will be discussed in the next section.

Following the standard thermodynamical definitions, the
magnetization per unit volume, the magnetic susceptibility and
the isothermal compressibility within the coexistence region can
be evaluated as
\begin{eqnarray}
\mathcal{M}&=&(1-\lambda)\,\mathcal{M}^a+ \lambda \,
\mathcal{M}^b,
\label{CRmagnt}\\
\chi_k&=&(1-\lambda)\,\chi_k^a+ \lambda \, \chi_k^b, \;\; k=1,2
\label{CRsusp} \\
K&=&\frac{(w^a-w^b)\left[\lambda K^b+(1-\lambda) K^a
\right]\,n^a\, n^b}{n\, \left[ w \,(n^b - n^a)+w^a n^a - w^b n^b
\right]}, \label{CRcomp}
\end{eqnarray}
where $n^{a,\, b}, \, w^{a,\, b}$ are the total density of
particles and the isospin asymmetry fraction in each phase.  The
partial contributions to  the magnetization $\mathcal{M}^{a, \,
b}$, the susceptibility $\chi_k^{a, \, b}$ and the compressibility
$K^{a, \, b}$ are evaluated in the same way as for a pure single
phase.

\section{Results and discussion}\label{SecIV}

For the Skyrme model the SLy4 parametrization is used, for which
$t_0=- 2488.91$ MeV fm$^3$, $t_1= 486.82$ MeV fm$^5$, $t_2=-
546.39$ MeV fm$^5$, $t_3=13777$ MeV fm$^{7/2}$, $x_0=0.834, \;
x_1=-0.344, \; x_2=-1, \; x_3=1.354, \; \sigma=1/6$
\cite{DOUCHIN}. The saturation density, binding energy,
incompressibility and symmetry energy are $n_0=0.159$ fm$^{-3}$,
$E_B=-15.97$ MeV, $K_0=229.9$ MeV and $E_S=32$ MeV, respectively.
Another significative quantity is the in-medium nucleon mass at
the saturation density,  for which $m^{\ast}/m=0.694$ is obtained.

In first place we discuss the effects of the Gibbs construction on
the pressure and the spin asymmetry coefficient $W/n$. These
results are shown in Fig.~1, for $B=10^{18}$~G and $T=5$~MeV,
which is representative for most of the cases studied in this
work. The Gibbs construction is shown in dashed lines and
replaces, within the coexistence region (CR), the plain results of
the model described in Section \ref{SecII}. In panel (a) it is
shown that the CR includes the instability region where the
pressure decreases with density. The Gibbs construction instead,
produces a rather linear increasing pressure. The density domain
of the CR is reduced and eventually vanishes, by increasing the
isospin asymmetry $w$, as well as the temperature (not shown in
this figure). For certain values of temperature and magnetic
intensity,  as for example $B=10^{18}$ G, $T=10$ MeV, the CR
disappears for asymmetries above a typical value $w_0 < 1$. In
these cases and for $w \simeq w_0$, a retrograde phase transition
takes place. This means that the transition starts and ends at
states of similar density and isospin composition, but in between
an admixture with states of very different conditions is
developed. We illustrate this phenomenon by including afterwards
the case $B=10^{18}$ G, $T=10$ MeV, $w=0.8$.

In the panel (b) of the same figure, we show the spin asymmetry.
It can be seen that protons and neutrons are highly polarized at
very low densities and they  depolarize progressively as the
density grows. The coexistence of phases induces an equilibrium
state with a significantly reduced degree of polarization, due to
the admixture with a higher density and weaker polarization state.
For higher magnetic intensities, such as $B=10^{19}$ G, the same
mechanism causes the frustration of the total neutron
magnetization, but it is not able to destroy the magnetic
saturation of the proton, as will be discussed subsequently.

The behavior of the pressure as a function of the density for
several temperatures and isospin asymmetries is shown in Fig. 2
for $B=10^{18}$ G and Fig. 3 for $B=10^{19}$ G. The points where a
sudden change of slope occurs, correspond to an endpoint of the
CR. They do not appear for some particular cases at $T=10$ MeV,
because the coexistence does not exist for $B=10^{19}$ G and
$w=0.8$ (Fig. 3b), whereas a retrograde transition goes on for
$w=0.8$, $B=10^{18}$ G (Fig. 2b), as explained above. For a given
magnetic intensity, thermal effects are more important for lower
densities. Furthermore, an increase of the magnetic intensity at
constant temperature induces an evident increment of the pressure
for $n/n_0 \lesssim 1$, but the opposite trend is observed for
$n/n_0 > 1$. It must be pointed out that the Gibbs construction
eliminates all the instabilities for the range of densities and
temperatures studied here. In particular there are no regions
where the pressure decreases with the density.

The density dependence of the spin asymmetry fraction is shown in
Fig. 4 for $T=5$, several values of $w$, $B=10^{18}$ G (Fig. 4a)
and $B=10^{19}$ G (Fig. 4b). Thermal variations are of no
relevance for this quantity. For the lower field intensity, the
proton relative polarization is enhanced as $w$ decreases, whereas
for the neutrons there is only a weak dependence on the isospin
composition. For $B=10^{19}$ G, the proton component is completely
polarized for all the range of $n$ and $w$. The effect of the
medium polarization is emphasized in this case, as can be seen in
the dependence on $w$ of the neutron spin asymmetry for $n/n_0 >
0.75$. For a fixed total density $n$, the neutron component is
completely polarized for $w=0$ and is progressively depolarized as
$w$ increases. This is a consequence of the dynamical coupling of
protons and neutrons (see Eqs.
(\ref{protSP})-(\ref{SkmPotential})).  It must be pointed out that
both components, but specially protons in a high $w$ sample, tend
to recover a high degree of polarization at densities $n/n_0
> 1.5$. This feature can be a manifestation of the abnormal
spontaneous magnetization described by the Skyrme model at extreme
densities.

Results for the free energy per volume $\cal{F}$ as a function of
the density, are shown in Fig.~5 ($B=10^{18}$ G) and Fig.~6
($B=10^{19}$ G). At relatively low densities the kinetic energy
and the repulsion between nucleons are small, while the effect of
the magnetic field becomes the dominant one. For medium and high
densities the repulsive character of the nucleon-nucleon
interaction and the kinetic energy dominate over the magnetic
field and the system increases its energy. This is clearly
depicted in Fig.~5, whereas for $B=10^{19}$~G in Fig.~6, only the
$w=0.8$ case fits this description. For other values of $w$ one
should go to higher densities to verify this behavior. From both
figures we can see that the addition of protons makes the system
to be more  bound and so does the increase of the magnetic field.
Thermal effects are weak, as opposite contributions tend to cancel
each other: the kinetic term increases with $T$, while the entropy
contribution does the opposite.

The magnetic susceptibility characterizes the response of the
system to the external field and gives a measure of the energy
required to produce a net spin alignment in the direction of the
field. We have found that this quantity at moderate field
intensities, is sensitive to thermal variations, hence we devote
Figs. 7-9 to give a more detailed description of the density
dependence of $\chi$. For all the cases shown, there is a low
density regime where the system has an almost saturated
magnetization (see Fig. 4), therefore the magnetic response is
nearly zero. As it was previously discussed, in most cases the
coexistence of phases frustrates the total magnetization and
consequently enhances the magnetic response within the CR. This
fact can be distinguished by an approximately linear increase,
with a small slope, of the susceptibility as a function of the
total density.\\
 The magnetic susceptibility shows a complex
dependence on the population of the Landau levels. Therefore, in
order to clarify the discussion afterwards, it is worth to make
some general considerations about this relationship. In first
place it must be pointed out that the population of the Landau
levels decreases with both $B$ and $w$, but it increases with both
$n$ and $T$. However, the distribution function becomes smoother
as the temperature grows, erasing eventually effects due to the
progressive occupation  of  different Landau levels. In second
place, when very high levels are occupied, further changes in the
quantum number $j$ has only imperceptible consequences on the
susceptibility. Finally, the proton component generally shows a
diamagnetic behavior, but it turns to be paramagnetic when the
statistical
occupation is peaked at the lowest Landau level  $j=0$.\\
Now we focus on the susceptibility for protons in the lower panel
of Fig. 7, where the case of $B=10^{18}$ G and $T=1$~MeV, is
illustrated. For increasing densities, the susceptibility changes
from a linear response to an oscillatory behavior. The linear
response turns out from the Gibbs construction, while the
oscillatory behavior is a consequence of the population of the
Landau levels. Note that in the evaluation of the partial
contributions $\chi_k^{a, \, b}$ in Eq.~(\ref{CRsusp}), the Landau
levels also play a role. However, the narrow range of variation of
the densities $n^{a}$ and $n^{b}$ allows the linear behavior for
the susceptibility. For $w=0$ and $w=0.4$ the system is
diamagnetic, while for $w=0.8$ it is mainly paramagnetic due to
the fact that only the lowest Landau levels are accessible. In the
upper panel of the same figure, the neutron susceptibility is
depicted. An important change in the scale is observed. The
neutron susceptibility is always paramagnetic and also has a
linear behavior for low densities which changes to oscillatory as
the density grows. Even though neutrons do not have discrete
levels, they are influenced by the dynamical coupling among
protons and neutrons (see Eqs. (\ref{protSP})-- (\ref{EffMass})).
For the case $w=0.8$ a sudden decrease at $n/n_{0} \approx 1.7$,
is observed due to a change in the spin configuration of the
system.

A small increase
of the temperature, keeping fixed the magnetic intensity, destroys
the oscillations at high densities. See Fig. 8 for $B=10^{18}$ G,
$T=5$ MeV. The magnetic response of protons becomes almost linear
for the full range of densities. Two ingredients contribute to
these results: the distribution function becomes smoother as
the temperature grows and the statistical occupation of the
Landau levels increase, being this last effect the dominant one.

For the two figures just described, there is a noticeable
difference of magnitude between the susceptibility of protons and
neutrons. The proton component is significantly more reactive to
changes of the magnetic intensity. The effects of changing the
magnetic intensity by keeping fixed the temperature are exposed in
Fig. 9, where we present the results for $B=10^{19}$ G, $T=5$ MeV.
A dramatic change of scale is apparent for both components, but
specially for protons. This fact is coherent with the saturation
of proton spin alignment (see Fig. 4b), as a consequence the
proton component experiences a change of regime from diamagnetic
to paramagnetic. In contrast, neutrons which only have a
paramagnetic channel, are completely blocked when the saturation
of spin alignment is reached. Hence we have $\chi_2 = 0$, as for
$w=0$. Note that for $B=10^{19}$~G, the statistical occupation of
the Landau level is strongly dominated by $j=0$, which explains
the paramagnetic behavior of the proton susceptibility.

In Figs.~10 and 11, the isothermal compressibility is presented as
a function of the density for $B=10^{18}$ G and $B=10^{19}$ G,
respectively. In both figures we show results for $T=1$, 5 and 10
MeV and isospin asymmetries $w=0$, 0.4 and 0.8. For all the cases
shown, there is a clear difference between the CR and the domain
of higher densities. In the last case, we found the typical
behavior of an almost incompressible fluid with a slightly
decreasing compressibility. Here the variation of temperature and
isospin composition have weak effects, but they become appreciable
for $B=10^{19}$ G. An exceptional behavior is found for
$B=10^{19}$, $w=0$, $T=1$~MeV, where a local maximum can be seen
around $n/n_0 \cong 1.9$. It seems to be a feature of the model
used and deserves a further investigation.\\
Within the CR, the compressibility decreases strongly for very low
densities and becomes increasing for moderately higher values. As
a consequence a local maximum of $K$ arises at the extreme point
of the CR. This description applies strictly to the lowest
temperature shown here, $T=1$ MeV. As $T$ is increased, the effect
becomes less pronounced, but is still perceptible for $B=10^{19}$
G, whereas it is completely missed for $B=10^{18}$ G.

As the last matter of analysis we present in Fig. 12 the phase
diagram for a fixed value $B=10^{19}$ G. Here the closed curves
correspond to the isothermal contour of the CR, for both the $P-y$
(Fig. 12a) and $n-y$ (Fig. 12b) planes. The CR for a given
temperature is enclosed by the corresponding curve. In the upper
panel for each curve there is a maximum value for the pressure, at the left and
below that pressure lies the lower density gaseous phase and, at
the right the higher density liquid phase. Above that pressure a
continuous passage from one phase to another occurs. As the
temperature grows the area within the contour is reduced and
eventually collapse to the critical point.  Note that at very low
pressures it is necessary to include states with a small proton
excess in order to complete the phase diagram.

\section{Conclusions}\label{Summary}

In this work we have addressed the properties of asymmetric
nuclear matter under the action of very strong
magnetic fields 
, temperatures below $10$~MeV and densities up to twice the
saturation density. For the nuclear interaction we have used the
non--relativistic Skyrme potential model (SLy4 parametrization)
within a Hartree--Fock approximation. We have paid special
attention to the low--density and low--temperature domain, where
the nuclear interaction gives rise to instabilities, commonly
associated with a liquid-gas phase transition.
If the Coulomb
force is not included, the equilibrium state of the system
separates spontaneously into coexisting phases. This phase
transition was studied in detail in the past~\cite{SCHRAMM, WEISE} and it has received
renewed attention recently in connection with
astrophysical investigations~\cite{GIULANI} and
also in heavy--ion collisions, where the liquid--gas instabilities
are related to the formation of fragments occurring
in finite nuclear systems~\cite{DAS}.

Here we introduce an external magnetic field as a new parameter
which modifies the properties of the coexisting phases. We propose
and analyze the spin polarization, the magnetic susceptibility and
the isothermal compressibility as physical quantities that reveal
the changes in the global configuration as
well as in the internal composition of the equilibrium state.
A related investigation, but restricted to neutron matter, has
been presented recently in \cite{UNLP2}. There, a comparison of
the predictions of very different models was made, which give us
some confidence about the common features and warn us about
possible singularities of the model chosen.

To obtain the physical equation of state, we have implemented the
so called Gibbs construction, which is the appropriate procedure
when there are multiple conserved charges.

Referring to the spin polarization, it is higher for very low
densities. Neutrons are in general less polarized than
protons. At extreme intensities ($B=10^{19}$ G) the effect of the
medium polarization is emphasized,  due to the dynamical coupling
of protons and neutrons (see Eqs.
(\ref{protSP})-(\ref{SkmPotential})) the relative magnetization of
the neutrons is enhanced by increasing the proton fraction. The
spin asymmetry depends weakly on the temperature, while it is
strongly affected by the magnetic field. The spontaneous
separation into independent phases reduces the degree of
polarization for both protons and neutrons. This can be
understood because for a given total density within the CR, one of
such phases has a greater partial density and a lower
polarization. This is the reason why neutrons do not reach the
magnetic saturation at
medium densities under the strongest intensity considered here.

The energy per volume shows
that the addition of protons makes the system more bound. And so
happens for increasing the magnetic field. At variance, the
dependence with temperature is rather small.

The magnetic susceptibility is the most sensitive quantity, as it
shows a strong dependence on the isospin asymmetry, the
temperature and the magnetic intensity. For protons these
dependencies manifest through the population of the Landau levels,
which also reflects in the neutron susceptibility due to the
dynamical coupling generated by the Skyrme model. The
susceptibility in the CR is almost linear.

The isothermal compressibility at very low temperature offers a
clear manifestation of the changes in the phase diagram. Out the
CR the results corresponds to an almost incompressible fluid,
decreasing slowly with density and showing a small dependence on
$w, \, T$ and $B$. Within the CR, the compressibility changes from
steeply decreasing at very low densities, to moderately
increasing. In such a way a local maximum appears at the end of
the CR. This behavior is attenuated by increasing the temperature.

An overview of the general phase diagram shows that the
coexistence of phases exists for all the range of magnetic
intensities of physical interest. The critical temperature lies
above $T=10$ MeV. An increase of the magnetic field has multiple
consequences, on one hand it produces an enhancement of the range
of pressures involved, while does not modify the density domain.
Furthermore the range of isospin asymmetries supported is reduced
and for the higher $w$ the system evolves through a retrograde
phase transition.

It is worthwhile to mention that most of this
conclusions become evident because we used an extreme magnetic
intensity. For weaker fields, effects such as the
temperature-polarization antagonism or the synergistic combination
of neutron excess and spin polarization, are still active but they
manifest diffusely. 

\section*{Acknowledgements} This work was partially supported by the
CONICET, Argentina. E. B. acknowledges the support given by the CONICET
under contract PIP 0032 and by the Agencia Nacional de Promociones
Cient\'{\i}ficas y T\'{e}cnicas, Argentina, under contract
PICT-2010-2688.

\newpage
\begin{figure}
\includegraphics[width=0.9\textwidth]{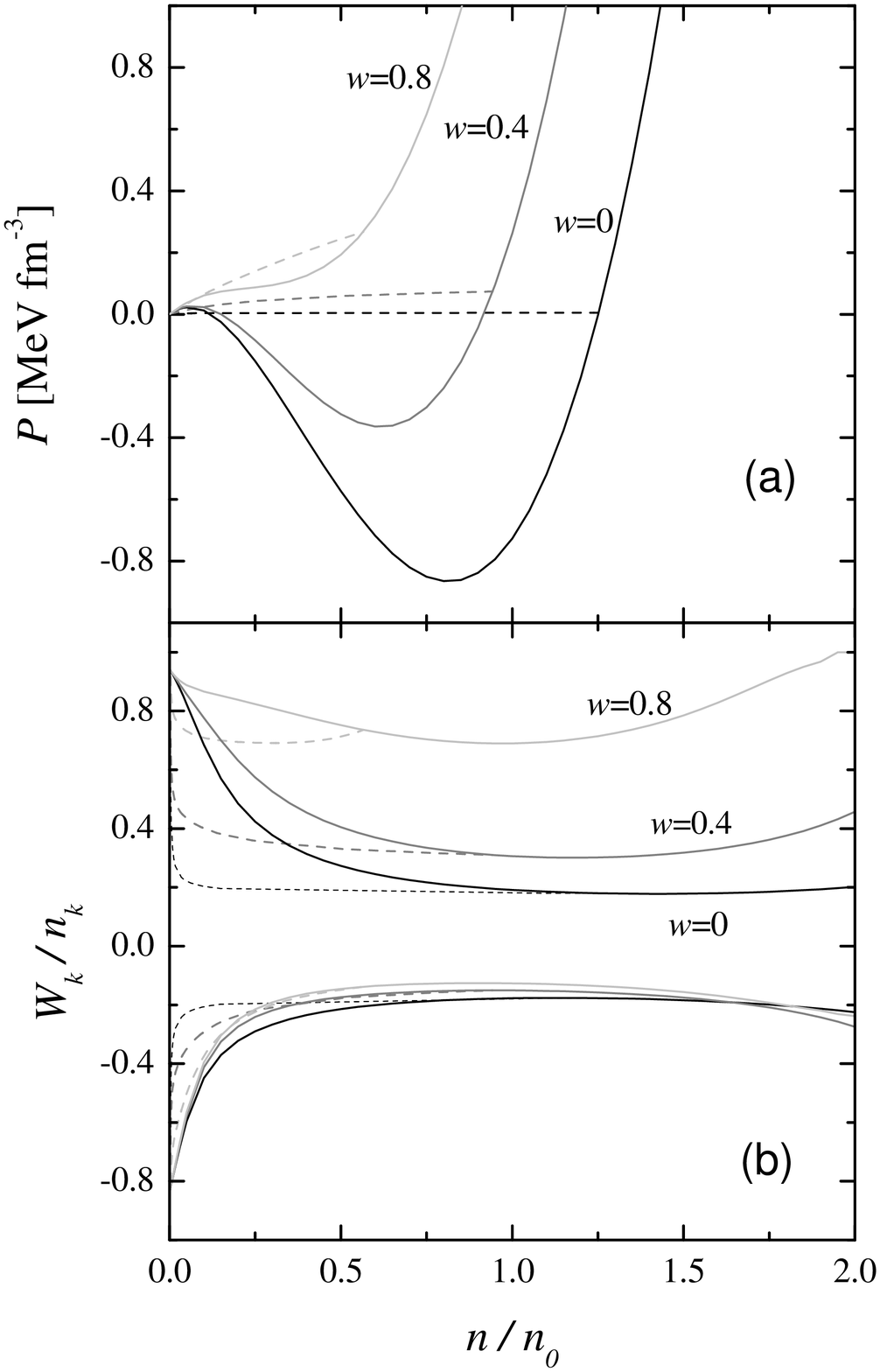}
 \caption{\footnotesize Details of the Gibbs construction
for $B=10^{18}$ G at $T=5$ MeV and several isospin asymmetries.
Panel (a) shows the pressure as a function of the density and
panel (b) the spin asymmetry fraction as a function of the
density. In the later case, the upper family of curves corresponds
to the proton. Dashed lines correspond to the Gibbs construction.}
\end{figure}

\newpage
\begin{figure}
\includegraphics[width=0.9\textwidth]{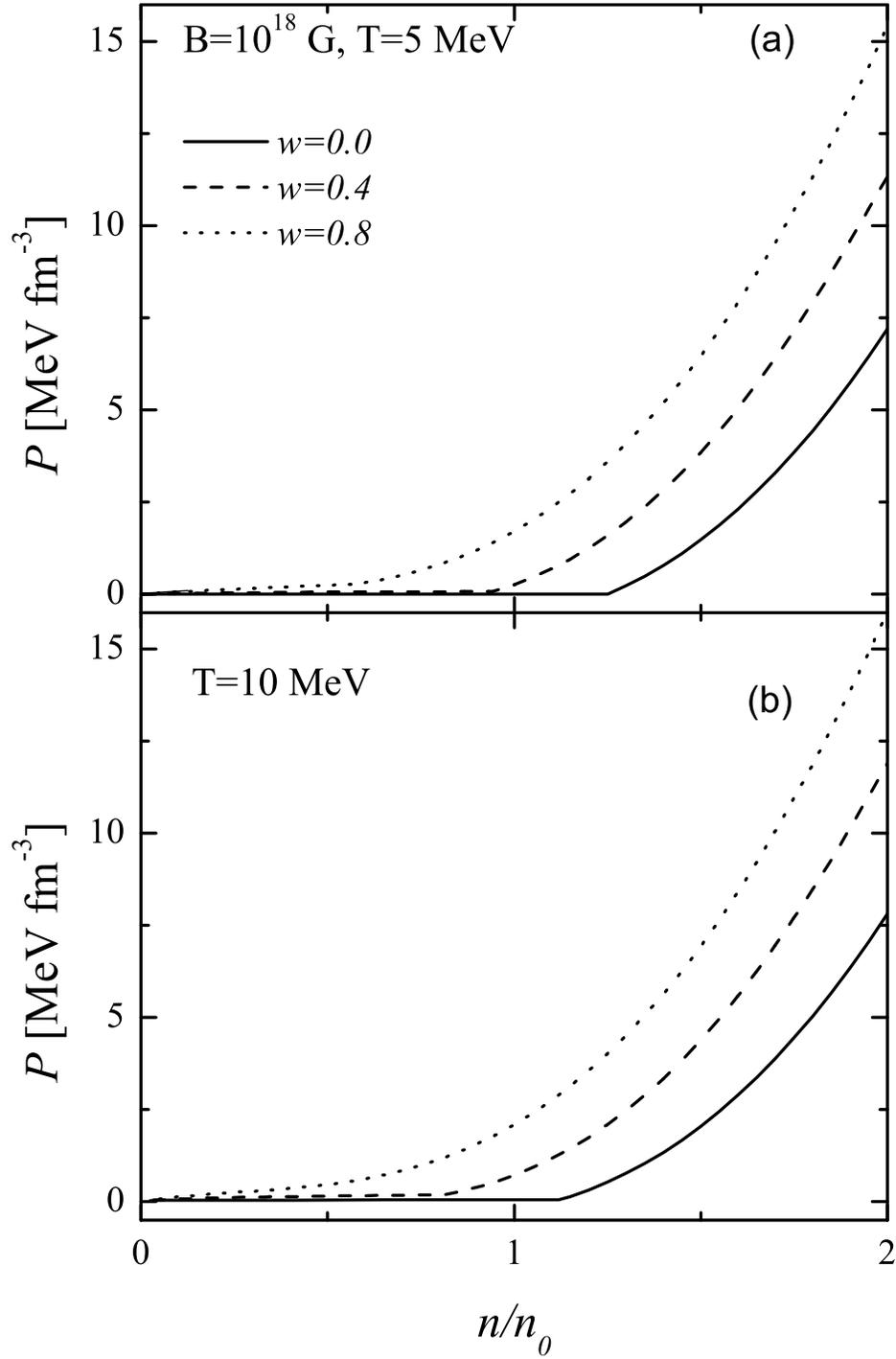}
\caption{\footnotesize Pressure as a function of the density for
$B=10^{18}$ G and several isospin fractions. Panel (a)
corresponds to $T=5$ MeV and panel (b) to $T=10$ MeV.}
\end{figure}

\newpage
\begin{figure}
\includegraphics[width=0.9\textwidth]{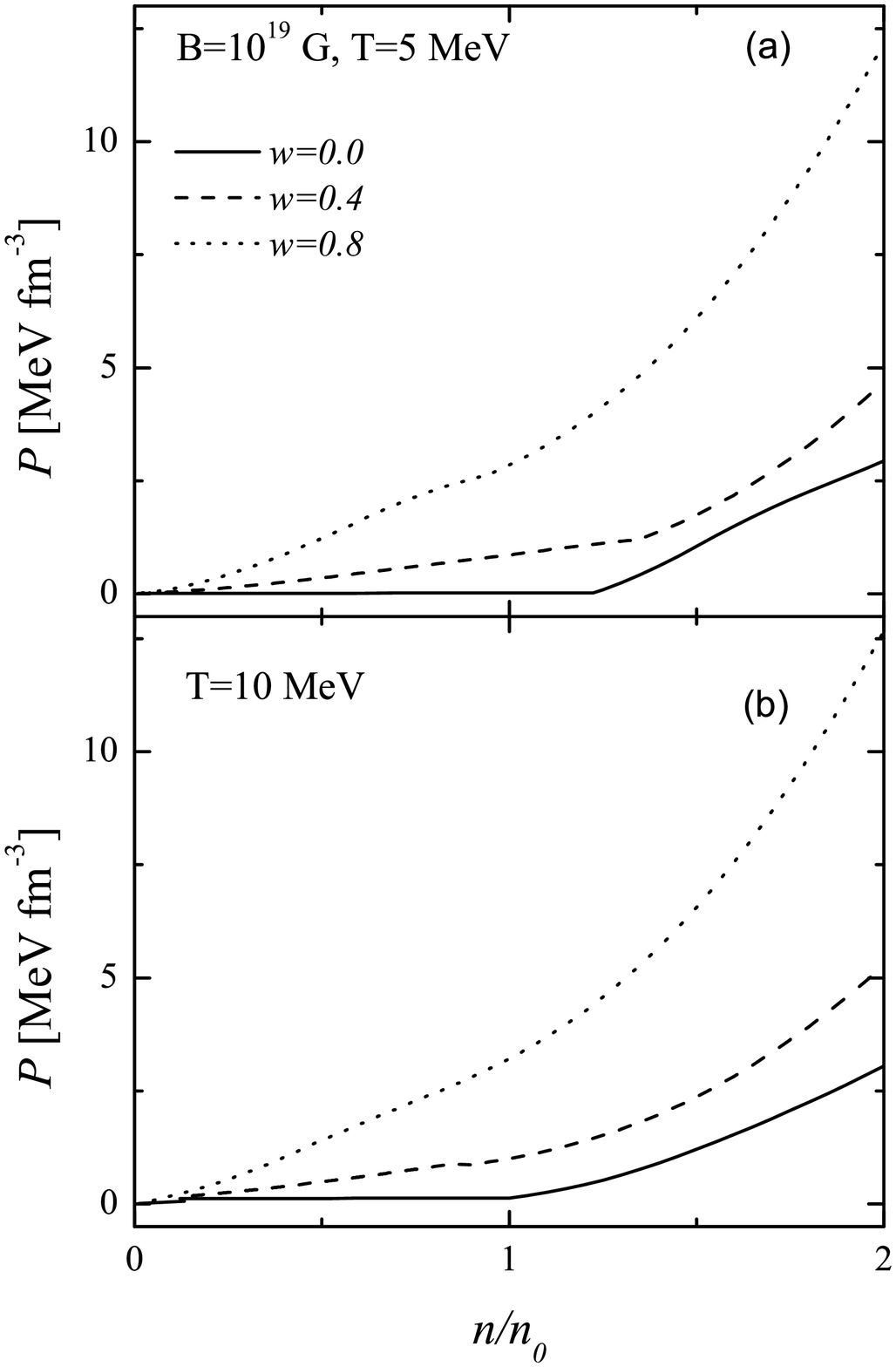}
\caption{\footnotesize The same as Fig.~2, but
for $B=10^{19}$ G.}
\end{figure}

\newpage
\begin{figure}
\includegraphics[width=0.9\textwidth]{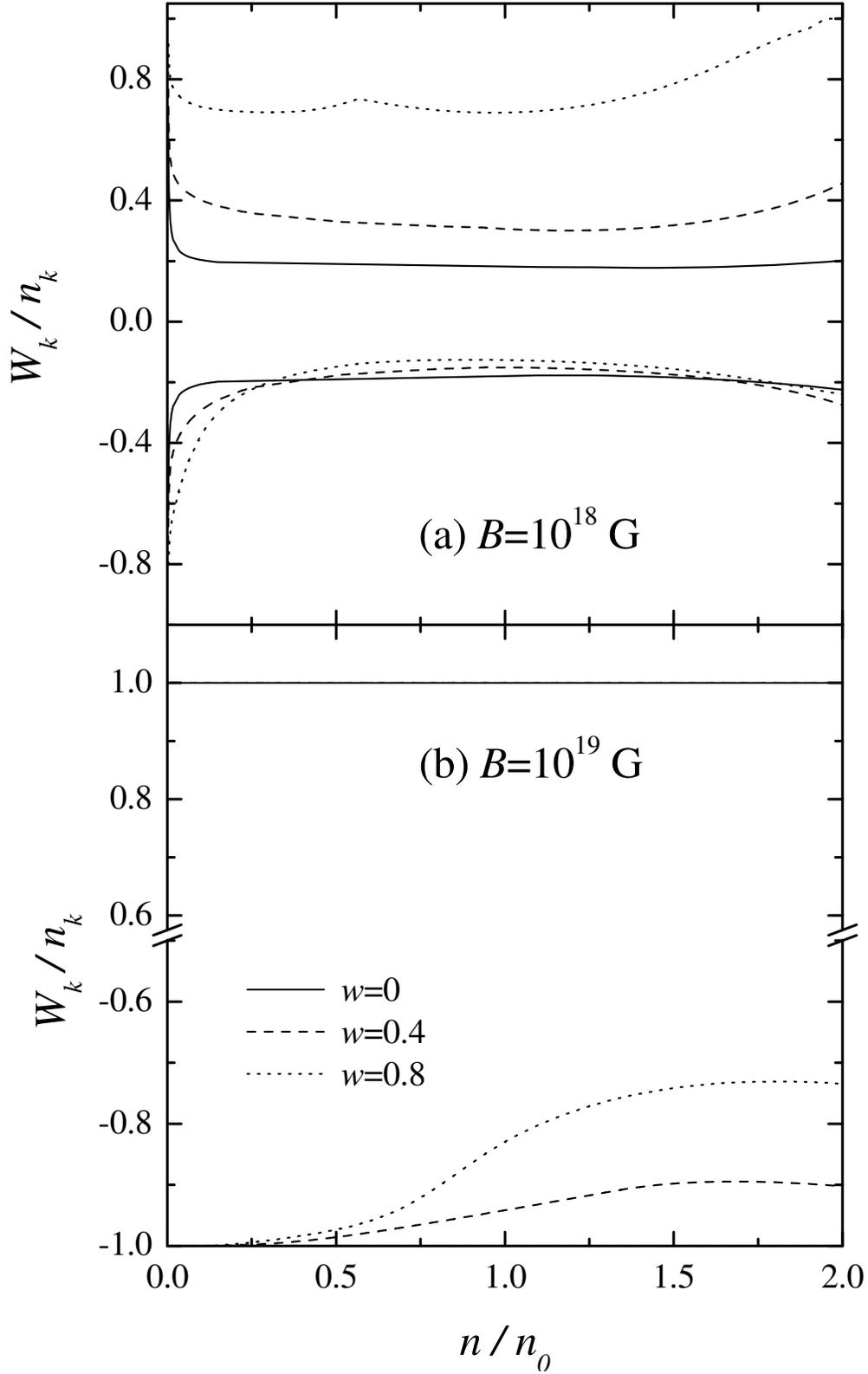}
\caption{\footnotesize The spin asymmetry fraction of protons and
neutrons as functions of the density at $T=5$ MeV and several
isospin fractions. Panel (a) corresponds to $B=10^{18}$ G and
panel (b) to $B=10^{19}$ G. In each panel the upper family of
curves correspond to protons. The same convention line is used for
both panels.}
\end{figure}

\newpage
\begin{figure}
\includegraphics[width=0.9\textwidth]{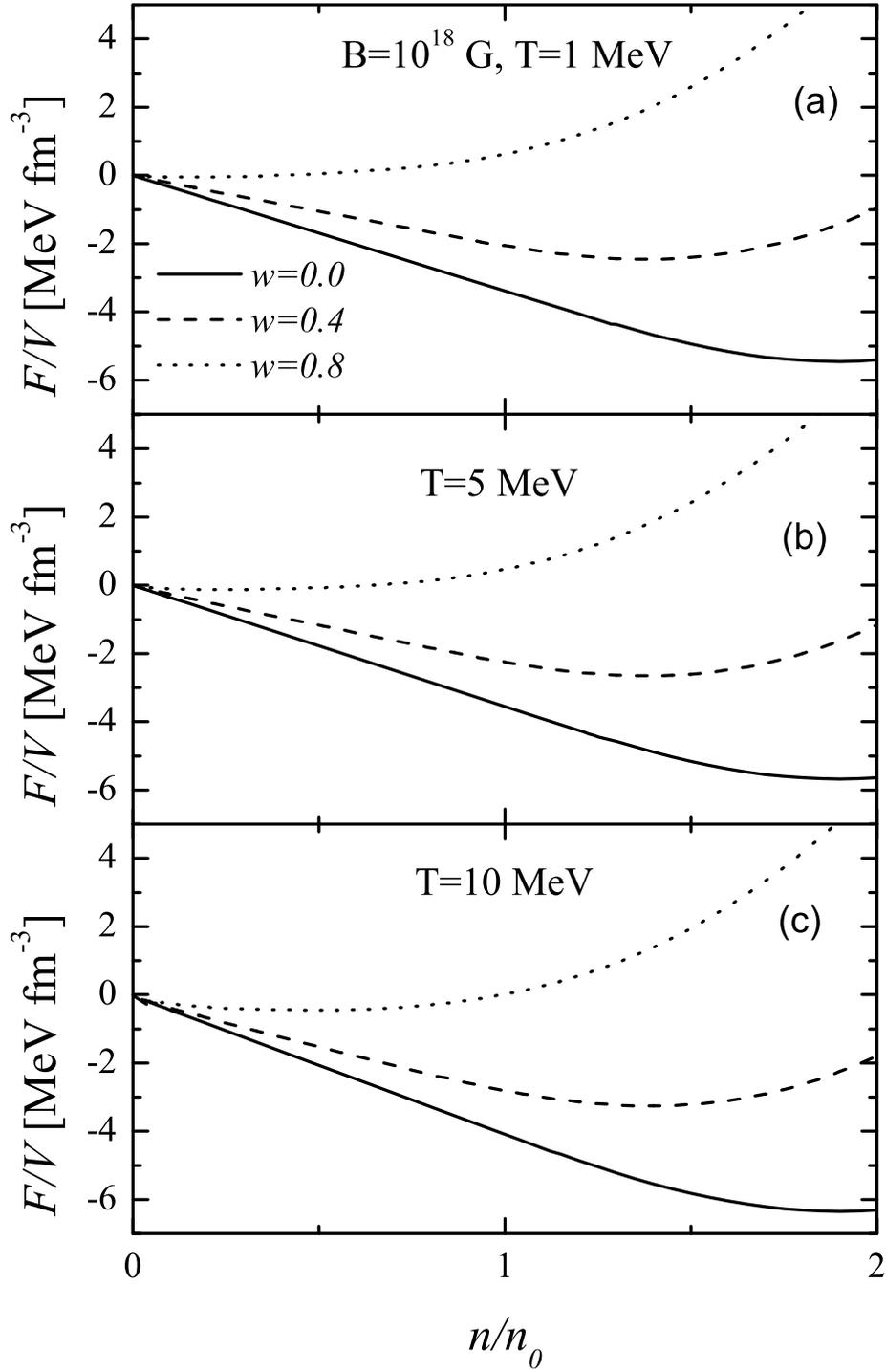}
\caption{\footnotesize The free energy per volume as function of
the density for $B=10^{18}$ G  and several isospin fractions.
Panel (a) corresponds to $T=1$ MeV , panel (b) to $T=5$ MeV and
panel (c) to $T=10$ MeV.  The same convention line is used for all
the panels.}
\end{figure}

\newpage
\begin{figure}
\includegraphics[width=0.9\textwidth]{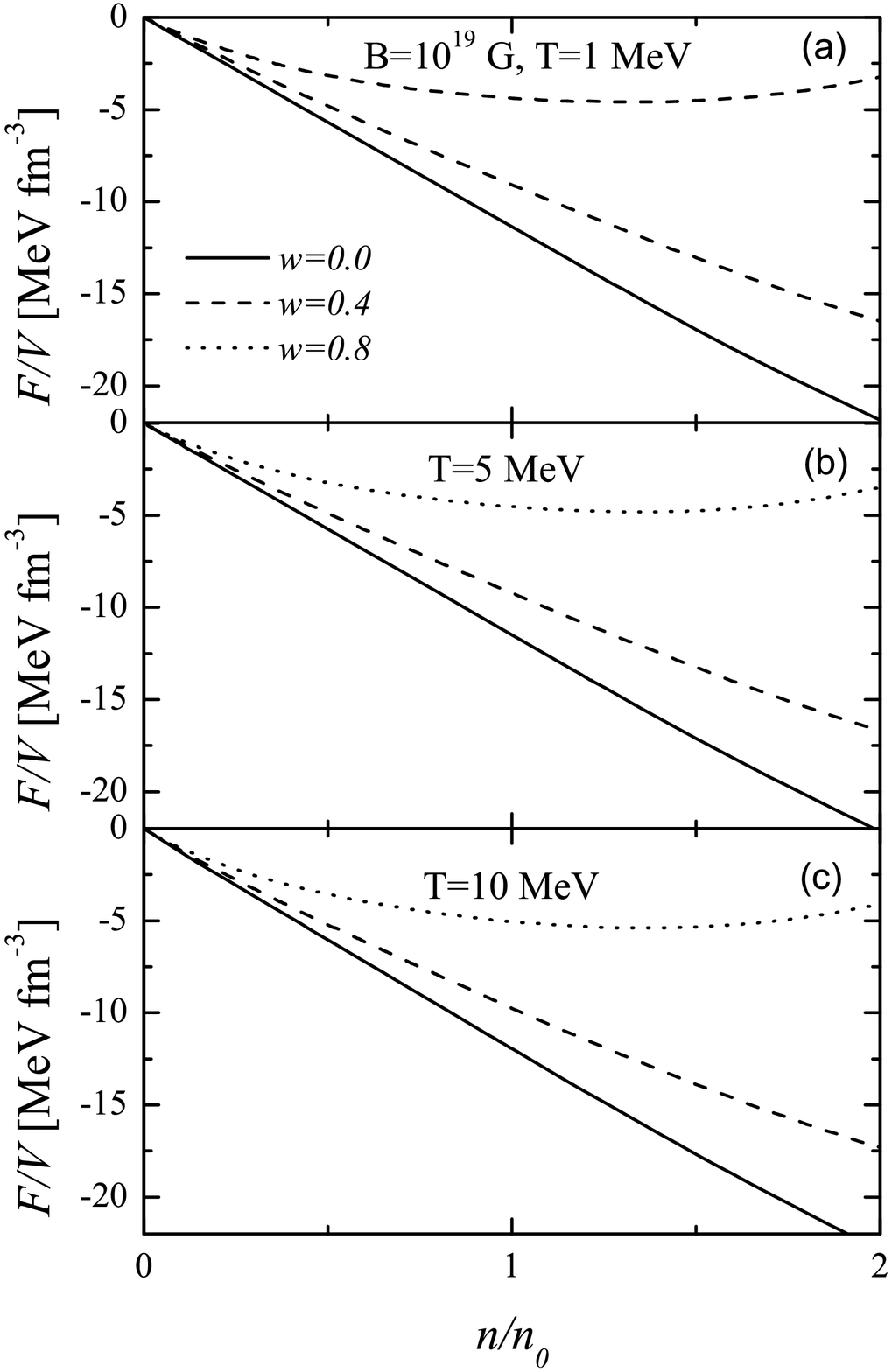}
\caption{\footnotesize The same as Fig.~5, but
for $B=10^{19}$ G.}
\end{figure}

\newpage
\begin{figure}
\includegraphics[width=0.9\textwidth]{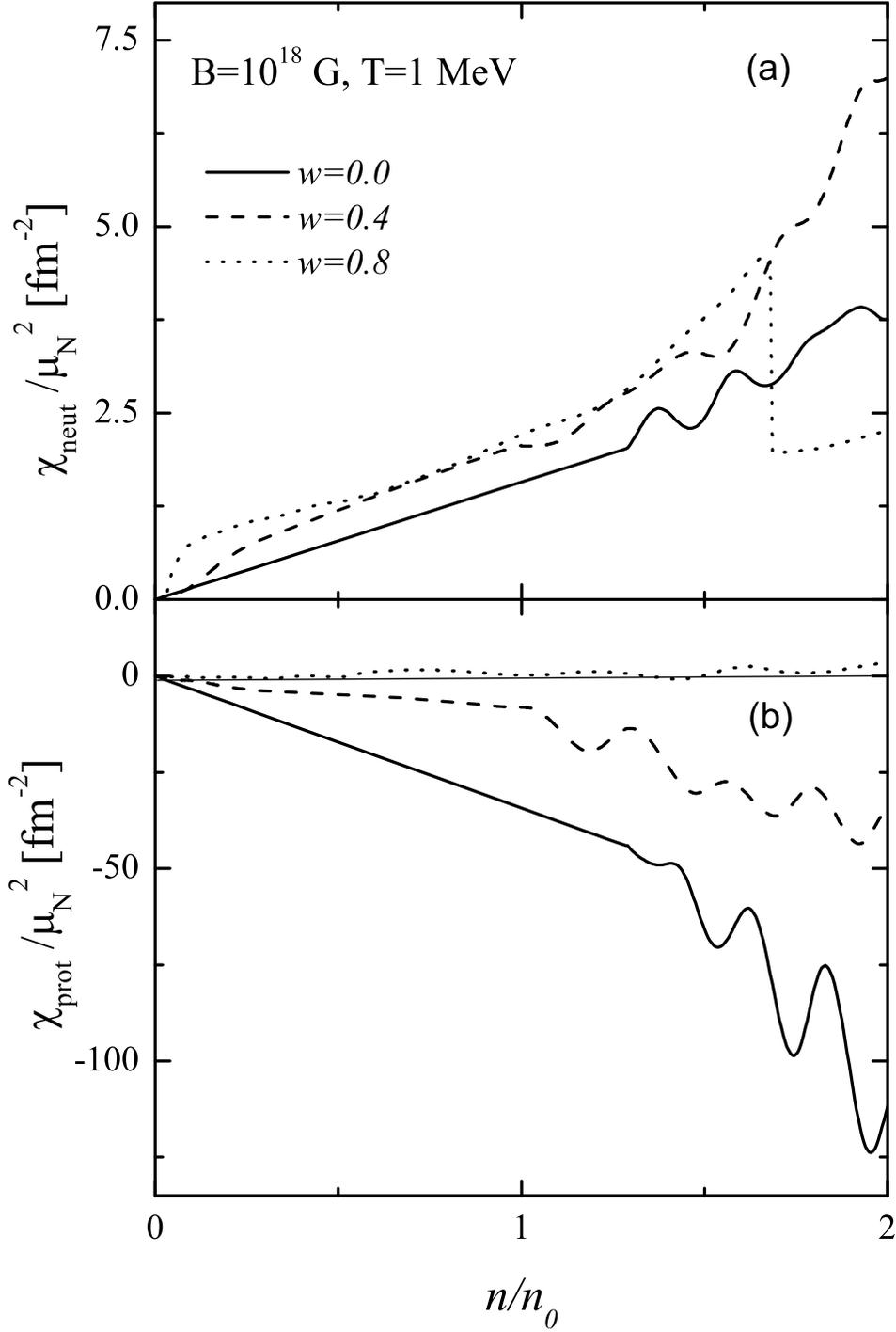}
\caption{\footnotesize The magnetic susceptibility (scaled with
$\mu_N^2$) as function of the density for $B=10^{18}$ G at $T=1$
MeV and several isospin fractions. Panel (a) corresponds to
neutrons and panel (b) to protons. The same convention line is
used for both panels.}
\end{figure}

\newpage
\begin{figure}
\includegraphics[width=0.9\textwidth]{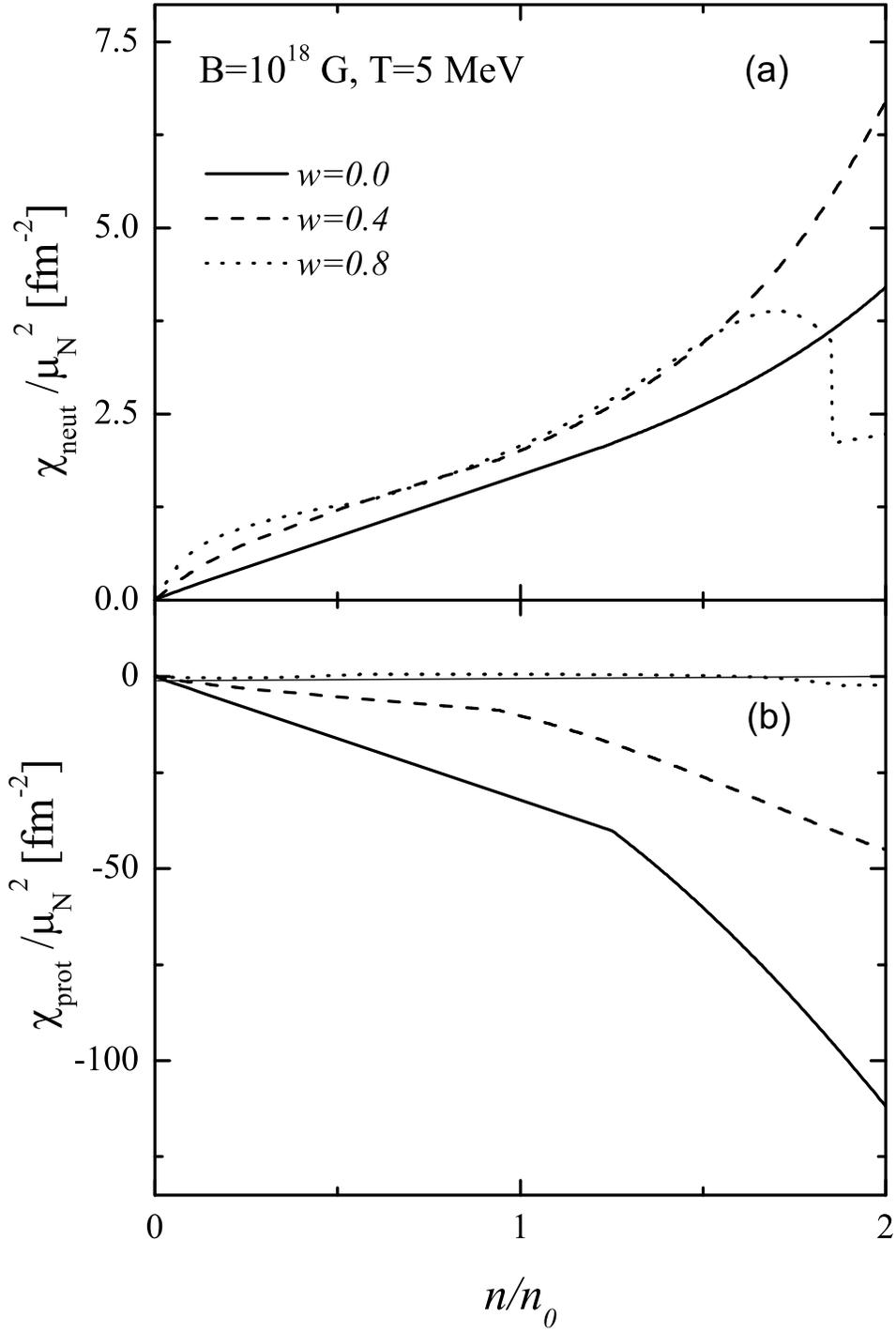}
\caption{\footnotesize The same as Fig.~7, but
for $T=5$~MeV.}
\end{figure}

\newpage
\begin{figure}
\includegraphics[width=0.9\textwidth]{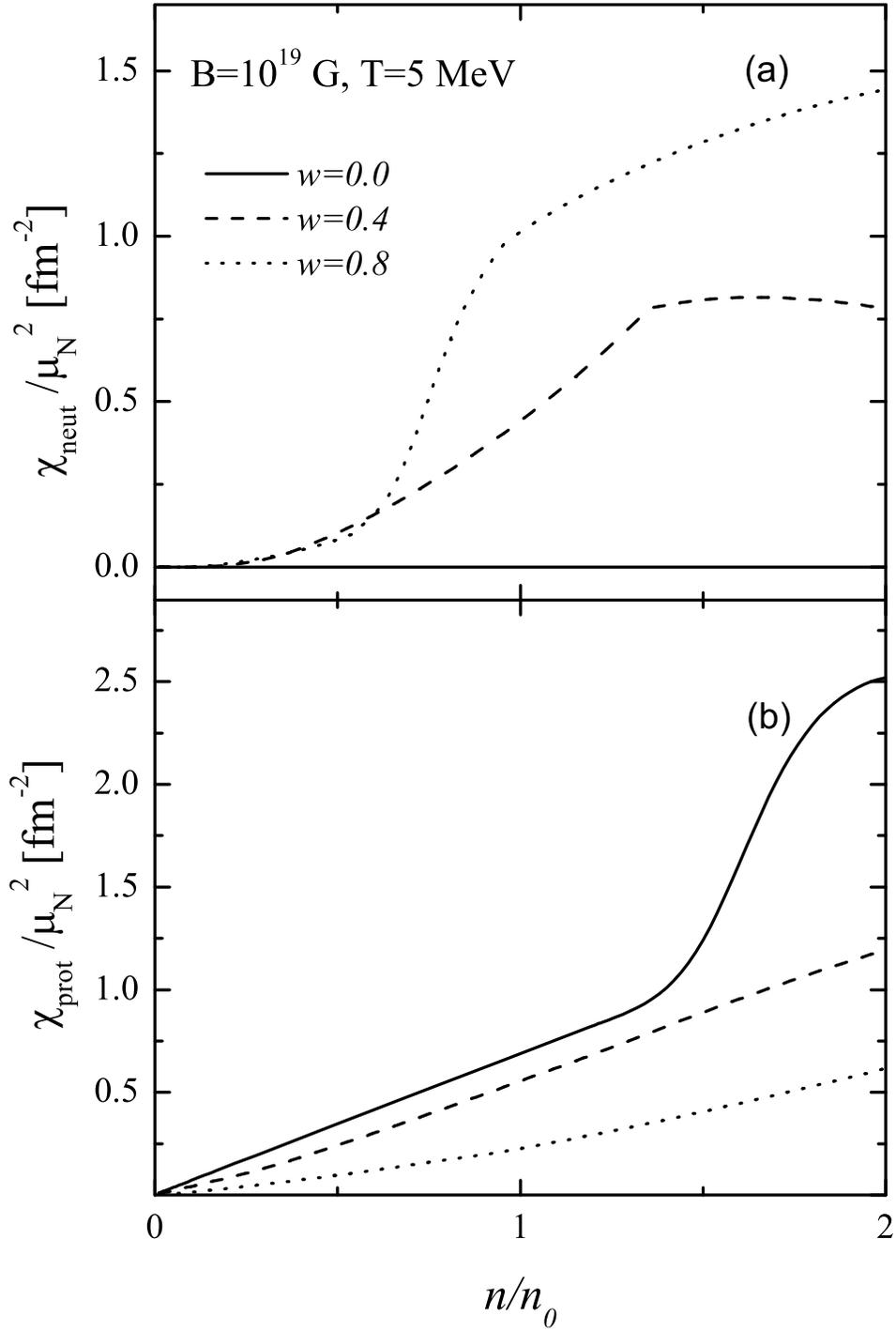}
\caption{\footnotesize The same as Fig.~7, but
for $B=10^{19}$ G and $T=5$~MeV.}
\end{figure}

\newpage
\begin{figure}
\includegraphics[width=0.9\textwidth]{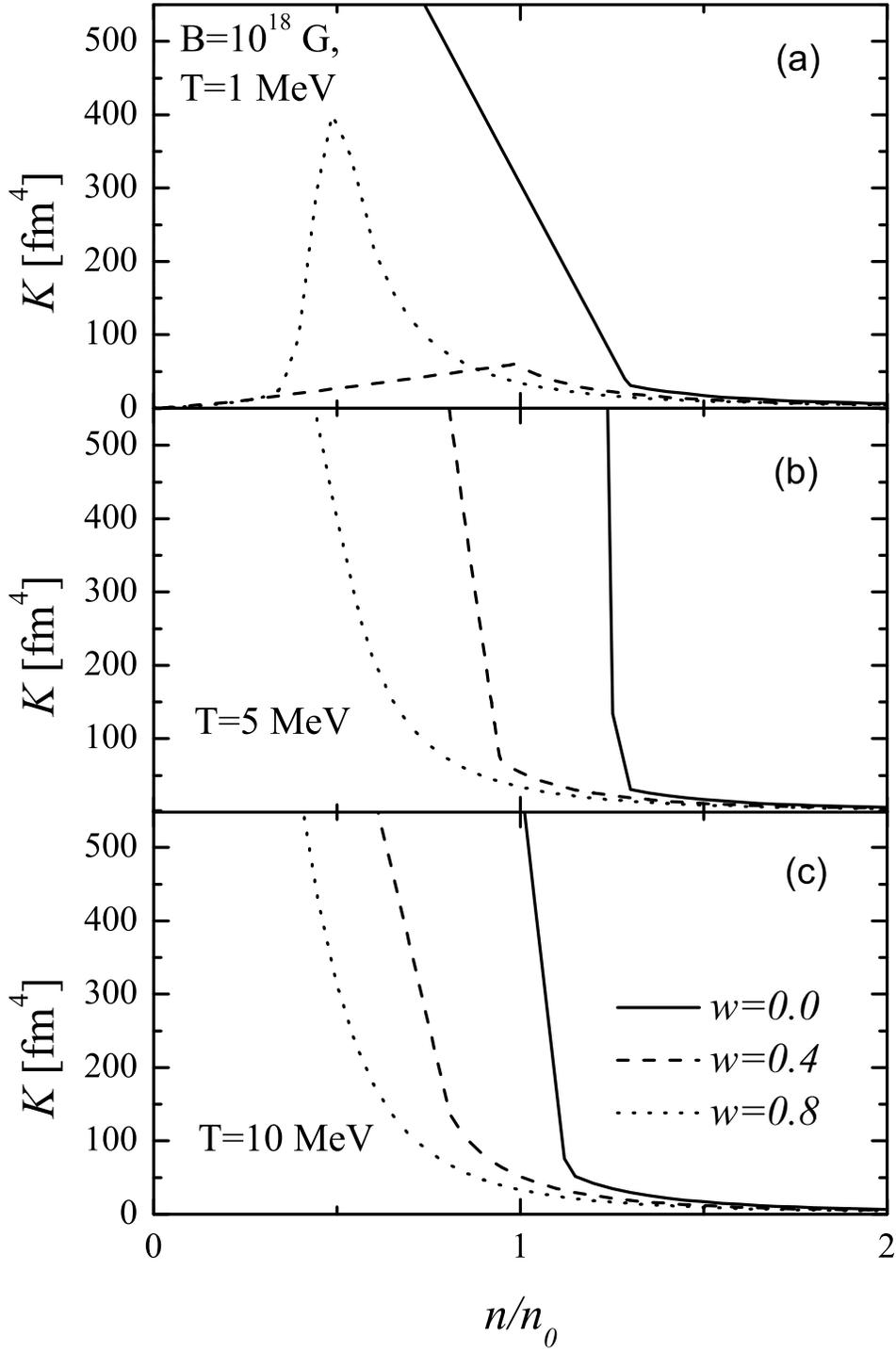}
\caption{\footnotesize The isothermal compressibility as function
of the density for $B=10^{18}$ G  and several isospin fractions.
Panel (a) corresponds to $T=1$ MeV , panel (b) to $T=5$ MeV and
panel (c) to $T=10$ MeV.  The same convention line is used for all
the panels.}
\end{figure}

\newpage
\begin{figure}
\includegraphics[width=0.9\textwidth]{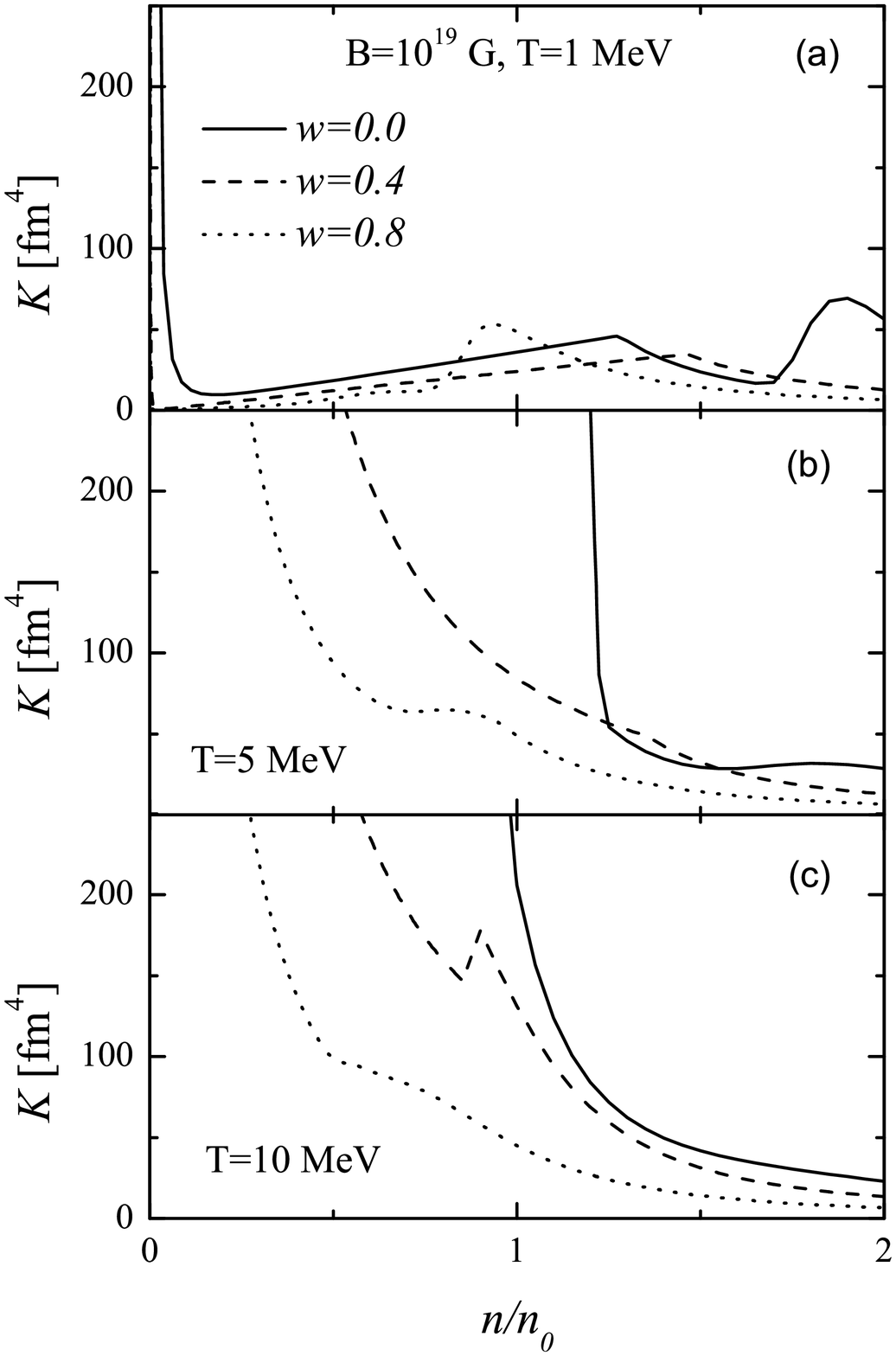}
\caption{\footnotesize The same as Fig.~10, but
for $B=10^{19}$ G.}
\end{figure}

\newpage
\begin{figure}
\includegraphics[width=0.9\textwidth]{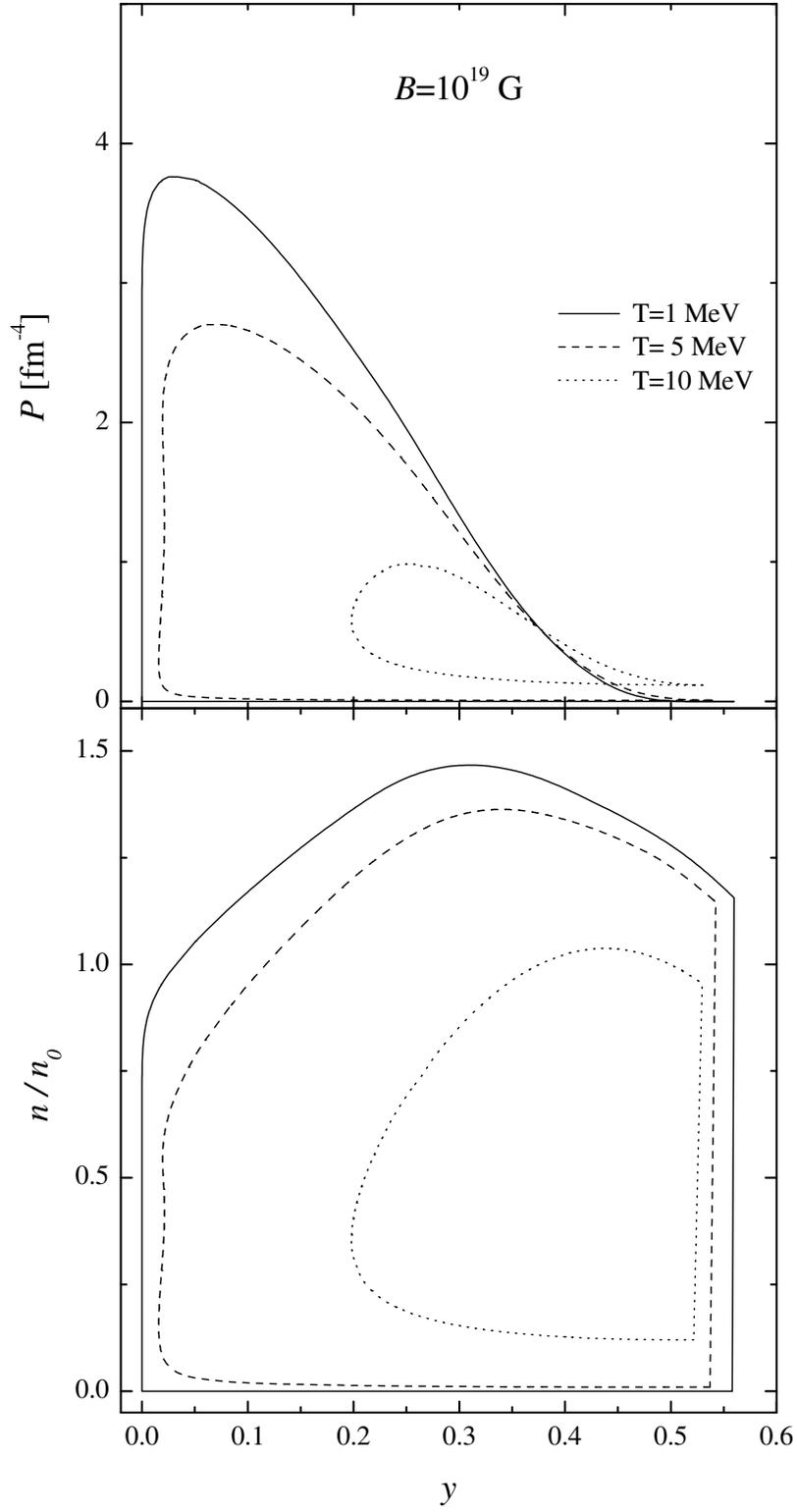}
\caption{\footnotesize The phase diagram for $B=10^{19}$ G and
several temperatures. Panel (a) corresponds to the pressure as a
function of the proton abundance  $y=n_1/n$, panel (b) to the
relative density $n/n_0$ as a function of $y$. The same convention
line is used for both the panels.}
\end{figure}

\end{document}